\documentclass{article}
\usepackage{graphicx} 

\title{How Do We Observe Relational Observables? }

\author{Emily Adlam  \thanks{Philosophy Department and Institute for Quantum Studies, Chapman University, Orange, CA92866, USA \texttt{eadlam90@gmail.com} }}
  \usepackage{natbib}
\date{\today} 
\setcitestyle{round}

\begin{document}

\maketitle

\begin{abstract} 

In theories with a diffeomorphism symmetry, such as general relativity and canonical quantum gravity, it is often proposed that the empirical content is encoded in relational observables. But how do relational observables actually make contact with experience?  I argue that this question can only be answered by providing a schematization of the observer which is appropriate for the context of a diffeomorphism-invariant theory. I suggest that this may require us to move away from a `passive awareness' conception of consciousness towards a more agential conception, because there is a clear sense in which an embodied agent must experience herself as localised at a time.  Finally, I  consider what this means for the prospects of using quantum reference frames to address the problem of time, arguing that the way in which quantum reference frames are standardly described does not give us adequate resources to model agency, so some other kind of internal reference frame may be required to fully explain how we observe relational observables  in the quantum context. 

\end{abstract}

\section{Introduction}

In both general relativity and canonical quantum gravity, there are some peculiarities around the notion of an observable: because of diffeomorphism invariance, the kinds of variables that would in a classical context encode the empirical content are not  observable in these theories.   An influential approach to this  issue employs relational observables to encode the empirical content. But despite impressive technical results in this area in recent years, the details about how these structures can make contact with experience remain somewhat unclear. So how exactly \emph{do} we observe relational observables?

Part of the difficulty is that observations culminate in conscious experience, and therefore telling a complete story about the process of observing a relational observable may require having something to say about the physical origins of consciousness, which is notoriously a  hard problem. Then of course, when we try to apply the relational observables approach in the quantum context, we run headfirst into the measurement problem. In this article I will try to make some headway on this difficult issue: I will take as a guiding principle the idea that consciousness can only supervene on or have access to facts which are physically real, and I will try to understand what kinds of facts could be suitable to play this role in the context of a diffeomorphism-invariant theory. 

A central part of the difficulty in both GR and quantum gravity is that relational observables are `timeless' in the sense that they have their values eternally and atemporally, which leads to a prima facie puzzle about how observers whose experience arises from  relational observables could still experience themselves as being located at specific times. I will argue that the solution to this problem may require us to move away from a `passive awareness' conception of consciousness towards a more agential conception, because there is a clear sense in which an embodied agent \emph{must} experience herself as localised at a single time. Thus I will argue that one way of making sense of experience in a diffeomorphism-invariant theory might involve postulating that consciousness supervenes on or has access to complete relational observables within the brain which represent some kind of  deliberative process. I will also argue that in the quantum gravitational context, it seems likely that the relations on which consciousness supervenes must be modal or nomic rather than spatiotemporal in character, which further supports an agential conception of consciousness. 

Finally, I will consider what this means for the prospects of using quantum reference frames to address the problem of time in the context of quantum gravity. I will argue that the way in which quantum reference frames are standardly described does not give us adequate resources to model agency, so if it is true that conscious experience is necessarily linked to agency, some other kind of internal reference frame may be required to fully explain how we observe relational observables  in the quantum context.

\section{Observables in Diffeomorphism-Invariant Theories \label{intro}}

Diffeomorphisms are a local symmetry of GR. This means that   given any history which is dynamically possible according to the equations of GR, we can take that history and apply a local diffeomorphism to the contents of some spacetime region, yielding another history with the same initial and final conditions which is also dynamically possible according to the equations of GR. So naively, it seems as though the theory fails to be deterministic: the initial and/or final conditions, together with the laws, are not enough to fully determine the history \citep{Wallacenew,Earman2002-EARTMM}.

Similar difficulties appear in other theories with a local symmetry group, which are often known as `gauge theories.' The typical way of handling  such  difficulties is to restore  determinism by simply stipulating that histories related by a symmetry transformation are  physically identical.  I will refer to the procedure of identifying symmetry-related histories as the `standard interpretation' of a gauge theory. In the case of GR, the standard interpretation ensures that the theory is invariant under diffeomorphisms \citep{Wallacenew,Earman2002-EARTMM}.

This has an important consequence:  variables which  `\emph{can be expressed as local functions of the coordinates}' \citep{rovelli2022philosophical} are clearly not invariant under diffeomorphisms, and thus  the standard interpretation of GR tells us that such variables are not only unpredictable by the theory, but physically meaningless. These quantities are not `observables' - and in the context of a gauge theory, this is understood to mean not merely that such observables cannot be observed by beings with our particular sensory faculties, but that they are not physically real at all.  As \cite{Einsteindialog} himself puts it, `\emph{the gravitational field at a certain place does not correspond to something `physically real,' but in connection with other data it does}'.
 
The  unreality of  variables pertaining to local functions of coordinates in a diffeomorphism-invariant theory leads to a prima facie puzzle, because at least na\"{i}vely it seems as though much of our ordinary experience of the world involves observing the values of locally defined variables at spacetime points or in spacetime regions - `\emph{all we can truly observe is localized – we have no access to infinity}' \citep{Gary_2007}. Yet surely such things cannot  feature in our experience if they are not physically real. So there are two related problems about diffeomorphism-invariant theories.   First, can we recover enough physically real structure in such theories to be able to formulate classical and quantum physics, in appropriate regimes? And second, can we understand how our experience makes contact with the physical world in such theories? I will refer to the former as the formal problem, and the latter as the experiential problem. Both  problems need to be solved if these theories are to be successful -   the formal problem because empirical adequacy requires that we can recover classical and quantum physics in some limit, and the experiential problem because empirical adequacy requires that  the theory can properly reproduce the observations which constitute the evidence for the theory, and those observations ultimately take the form of experiences.  As \cite{Healey2002-HEACPC} puts it, `\emph{The evidence for any physical theory is empirical: it consists, ultimately, in the results of observations and experiments. Whatever physical form these take, they must give rise to experiences in scientists who perform them if they are to serve their epistemic purpose.}'

To be clear, in emphasizing the experiential problem it is not my intention to suggest that either experience or consciousness must play any special role in the relational observables framework, or indeed in quantum mechanics. Quite the reverse: it has \emph{always} been true that in order for our theories to have empirical content, they must be able to tell us how the relevant theoretical structures can be related to experiences. In the terms used by \cite{curiel2020schematizing} and \cite{Stein}, a physical theory which has some claim to apply to our actual world must include a `schematiziation of the observer,'  which `\emph{identifies the junctions where meaningful connections can be made between the (theory and experience) and embodies the possibility of the epistemic warrant we think we construct for our theories from such contact and connection}' \citep{curiel2020schematizing}. Without such a thing, we would have no way to understand how the theory is connected to the observations that are supposed to be the evidence for it, so the theory would not be able to be empirically confirmed by  those observations.

Of course, this does not mean that a scientific theory should be expected to solve the hard problem of consciousness - to connect a theory to the relevant observations it is usually enough to identify in approximate terms the kinds of variables which consciousness may supervene on or have access to, without saying much about consciousness itself. For example, in the classical picture it is natural to assume that consciousness supervenes on or has access to something like instantaneous states of the brain, so in that context, we can   understand how the theory is connected to the observations which are the evidence for it if we are able to imagine some kind of physical mechanism by which the state of the brain can become correlated with the kinds of facts about external reality which are supposed to be described by the theory. 

The relational observables scenario, then, is not really different in principle from the ordinary classical case. Its novelty is simply that some of the types of variables which we might have expected consciousness to supervene on or have access to in a classical context, such as instantaneous states of the brain at various times, are not physically real in a diffeomorphism invariant theory, and thus  in order to connect up the theory with the observations we will have to identify some \emph{other} class of variables which consciousness could supervene on or have access to. Thus in the relational observables scenario we may   have to tell a story about observation which is a little different from the one we might have told in the classical context.

Why does this matter? First of all, it is important from the point of view of epistemic rationality - if we cannot offer any coherent story about how our theories are related  to the observations which are our evidence for them, then we cannot claim that our epistemic commitments to such theories are well-justified. But second, there are a number of unresolved conceptual issues in modern physics which could could potentially be related to inadequacies in our understanding of observation or experience. For example,   the quantum measurement problem is at least partly about how it is that we come to experience unique measurement outcomes, and the problem of time is at least partly about how we come to have temporal experiences. So it is possible that some of our difficulties in this vicinity arise from misconceptions or false assumptions about the way in which our conscious experience makes contact with physical reality. To be clear, I am not suggesting that solving these problems will require us to treat consciousness as fundamental or nonphysical - I am simply noting that our understanding of the relation between conscious experience and the physical world may have some bearing on how we approach these questions. Of course there is no way to know in advance whether or not this line of enquiry will be productive, but given the persistence of these problems it seems like a good idea to critically examine all concepts which could be relevant to the difficulties.

\subsection{Identifying Observables}

So how are we to go about solving the formal and/or experiential problem? 
In both cases, our starting point must presumably be to identify some variables which \emph{are} observables. In a diffeomorphism-invariant theory, there are two main classes of such variables: `\emph{highly nonlocal quantities defined over the whole spacetime  and (differently) non-local, `relational’ quantities built out of correlations between field values}' \citep{pittphilsci4223}.  In this article I will focus on approaches which make use of the second class of variables, which are often known as  relational observables. The basic idea of these approaches is that `\emph{localization of an observation must be performed in relation to some background state (us, our detector, the planet, etc.)}'  \citep{Gary_2007}, and thus it is really relational observables, rather than local functions of coordinates, that we ultimately observe. 

I will discuss  two specific formalisms seeking to implement the idea of relational observables: partial observables, and quantum reference frames.  Both of these formalisms seem to be  primarily intended as solutions to the formal problem, which is understandable, since the formal problem is much easier to mathematize and does not require us to confront thorny philosophical questions about the nature of consciousness. But can these formalisms also solve the experiential problem? In order to answer that question, we must try to understand how it is that we actually come to observe these variables - what does the process of observation involve in a diffeomorphism-invariant picture?

It should be noted that these concerns around observation are also likely to apply to quantum gravity, since many physicists expect that a full theory of quantum gravity will also obey diffeomorphism invariance. Indeed, the mystery around diffeomorphism invariance becomes deeper when we move to quantum gravity, because the canonical quantization procedure developed by Dirac apparently forces us to understand the procedure of identifying transformations related by a gauge transformation as telling us  `\emph{that certain configurations are really the same configuration}' whereas in classical GR we have the option of understanding it as telling us `\emph{that the same history can be described by many different sequences of configurations}' \citep{Wallacenew}. The former interpretation is puzzling on a conceptual level, since it seems to entail that time evolution does not produce any real change - quantum gravity is in some sense `timeless,' and this difficulty has become known as the `problem of time.'   Proponents of relational variables typically argue that the problem of time can be solved by recognising that even in classical GR it is only relational variables that are truly observable, which means that `time' can only be understood in terms of a correlation between a physical system and a clock; thus, once we understand   classical GR correctly, we see that the problem of time is not really a problem, since we should never have expected `time' to act as an external parameter in the first place\citep{rovelli2021layers}. 
But an important prerequisite for using relational variables to address the problem of time is that we must be able to connect relational variables up to experiences, so addressing the experiential problem is in some sense a necessary condition for fully resolving the problem of time.  I think it is only a necessary and not a sufficient condition, so I will not attempt in this article to assess whether the relational approach can \emph{fully} resolve the problem of time, but I do believe that achieving clarity on how relational observables are supposed to solve the  experiential problem may ultimately contribute to the larger project of solving the problem of time in quantum gravity.

\subsection{Partial Observables}

Rovelli defines partial observables as physical quantities `\emph{to which we can associate a measuring procedure leading to a number}' \citep{Rovelli_2002}. Meanwhile, he uses the term `complete observables' to refer to physical quantities which can be predicted, either probabilistically or deterministically,  by the relevant theory  - i.e. complete observables are simply those quantities which are usually called `observables' in standard gauge theory terminology. Rovelli argues that while partial observables themselves are not observable in the gauge theory sense, the \emph{relations}  between partial observables are complete observables - they form a special subset of the complete observables, which we might call the complete relational observables.   As a motivating example, Rovelli notes that when predicting the motion of a pendulum, we don't just predict velocities on their own - rather we make relational predictions which can  be expressed in the form `the velocity of the pendulum is $v$ at the time when the clock is reading $t$.' In this example, the velocity of the pendulum and the reading of the clock are both partial observables, but the relation between them is a complete relational observable. So as \cite{pittphilsci4223} puts it, `\emph{the dynamics is then spelt out in terms of relations between partial observables. Hence, the theory formulated in this way describes relative evolution of (gauge variant) variables as functions of each other,}'

Partial observables coordinatize an extended configuration space, while complete observables coordinatize an associated reduced phase space \citep{pittphilsci4223}. The space associated with the partial observables thus has additional degrees of freedom  - for example, in the case of a diffeomorphism-invariant theory, the extra degrees of freedom will encode the locations of fields or physical objects on the underlying manifold. As explained by \cite{Rovelli_2014}, although the extra degrees of freedom used to express partial observables are not diffeomorphism-invariant, nonetheless they are physically significant because they represent `\emph{handles through which systems can couple.}' That is, when we bring two systems together, the resulting  set of complete observables is not just the union of their individual complete observables - we also get some new complete observables describing relations between the systems, which can be characterized as relations between their partial observables. Thus the extra degrees of freedom in the extended configuration space are  physically meaningful since they provide a representation of the  possible ways in which a given system can form new complete observables with respect to some other system.

Note that as described above,  part of the rationale for the use of gauge theories  is the idea that when a theory is agnostic between two or more possibilities there is no real physical difference between those possibilities.   Thus although the space in which  partial observables are expressed has meaningful physical content in describing possible couplings, the standard interpretation of a guage theory tells us that the partial observables themselves cannot   be physically real - they are not individually predictable by GR, so they are meaningless without a specification of what they are relative to. Some authors interpret Rovelli as denying the standard interpretation and asserting that partial observables are in fact independently real, but I will argue in section \ref{partial} that there is another way of reading Rovelli which does not  have this consequence.

\subsection{Quantum Reference Frames}

There are a number of different research programmes focusing on the general issue of  reference frames in quantum mechanics; in this article I will focus on the quantum reference frame (QRF) formalism as studied in works such as \cite{delahamette2021perspectiveneutral,article3,2020qctloe,2020acop,2020htsbrqc,giacomini2021quantum}. In these approaches, we define relational variables by choosing a subsystem of the universe   and then using it as a reference frame\footnote{The QRF formalism as discussed here is distinct from the study of reference frames in quantum information  \citep{RevModPhys.79.555} and the operational approach \citep{carette2023operational} although there are of course some relations between these various approaches.}.   

For example, one important special case  involves choosing as a reference frame a subsystem which can be regarded as a clock, so we can then write down the state of the rest of the world relative to the clock showing a certain time reading. This special case is similar to the framework developed by Page-Wootters for the purpose of addressing the problem of time in covariant quantum gravity, and in that formalism it can be shown that  relative to an `ideal clock' for which clock states are orthogonal, a quantum system will undergo the standard Schr\"{o}dinger evolution \citep{2020cqtd}, so it appears that by this kind of method we can recover ordinary quantum physics. 

An important consequence of the QRF formalism, as discussed by ref \cite{article3}, is that superposition is  reference-frame dependent.   For example suppose you are observing a quantum particle which is in a superposition of the position basis states $|0 \rangle$ and $|1 \rangle$. But that particle is only in the superposition relative to your reference frame - the QRF formalism provides transformations which we can use to switch into the reference frame of the particle instead, and in that reference frame \emph{you} are in a superposition in the position basis!

\section{Observing Partial Observables \label{partial}}

In this section I will seek to understand how it could  be physically possible to observe relational variables. For now I will work in the context of  classical GR,  deferring specific issues that appear in the quantum context until section \ref{quantum}. 

On the face of it the partial observables approach appears to offer a straightforward answer for the question of how we come to observe relational variables: we do it by observing partial observables, which by stipulation are those to which we can associate a measuring procedure. However, how exactly this works remains quite hazy.   For example, \cite{Rovelli_2002} offers the following  heuristic description: `\emph{Suppose we are in a (very simple) laboratory, and we want to check the correctness of (the equation of motion for a pendulum) ... Clearly we need two measuring instruments: one that gives us the pendulum position q and one that gives us the time t. The theory cannot predict the value of t. Nor can it predict the value of q, unless we specify that the value of q we are interested in is the one at a certain given time.}'  This description seems to suggest that we are able to measure the complete relational observable by first measuring the partial observables $q$ and $t$ separately and then simply comparing the values. 

But there is something a little puzzling here. For example, in the case of a complete relational observable which takes the form of a spatiotemporal coincidence between a clock variable and the position of an oscillator, \cite{Earman2002-EARTMM} argues that    `\emph{the measuring procedure cannot work ... by separately measuring the values of the clock variable and the oscillator position and then checking for the coincidence. For the positions of the clock and the oscillator are gauge dependent quantities ... Rather the measurement procedure must be directly responsive to the coincidence of values itself, even though the coincidence is not a coincidence of the values of observable quantities.}' That is, if it is true that partial observables on their own are not elements of reality,  we surely cannot measure complete relational observables by measuring two partial observables separately and then comparing their values; we must somehow measure the  complete relational observable all together.
 
Now, one might be tempted to respond to this concern with scepticism similar to that expressed by \cite{Maudlin2002-MAUTMM}, who imagines seeking to measure the value of a quantity at the spacetime point where two  geodesics coincide, and  argues that `\emph{one would not tell where the geodesics coincide in anything like (the way that Earman describes): one would tell by sending a rocket along each path and making the measurement when they collide. Nothing in any of Earman's arguments suggests any difficulty about this procedure.}' And of course, Maudlin is quite right that we all know  this operational procedure is possible, and that it is an effective way of making an observation. But this doesn't really address Earman's concern, because in this procedure it still appears that we need to observe two separate facts and then compare them: first, the  rockets crossed at the spacetime point $X_0$, and second, the result of our measurement was $O$ at the nearby spacetime point $X_1$. Expressed thus, these facts are not complete observables, so we should not be able to observe them separately and then compare them.

Perhaps Maudlin would argue that what we are really observing in this case is simply a point-coincidence, i.e. we observe that the rockets cross at the very same point as the measurement is made. But these events cannot  happen at \emph{literally} one and the same spacetime point, since the rockets and the measuring device cannot occupy the same point at the same time. Thus these events must simply happen in close proximity, in which case the difficulty described above still applies - one might  think that in order to determine that the events occur in close proximity, we would need to take note of the spacetime location of each event individually and then make a comparison to see that their spacetime locations are close, and yet we surely cannot do this if the locations are not individually physically real.  Obviously we are somehow able to make the observation nonetheless -  but \emph{how} do we do it? 
 
One option here is to read Rovelli as simply asserting,  in opposition to the standard interpretation of a gauge theory, that partial observables are in fact physically real and hence measurable, despite being unpredictable by GR. For example, \cite{pittphilsci4223} writes, `\emph{Both spaces - the space of genuine (complete) observables and partial observables—are invested with physicality by Rovelli; the partial observables, in particular, are taken to be physical variables.}'  However, there is an alternative possible reading of Rovelli based on the observation that observers are, themselves, parts of the physical universe. For Earman's worry as expressed  above seems to envision an observer external to  the universe,  measuring the clock variable and oscillator position and then comparing them; but of course in practice observers are physically embodied and thus they stand in physical relations to the variables they are attempting to measure, which provides us with some additional resources to draw on. So a natural way to think about Rovelli's proposal is to say that an observer is able  to observe partial observables relativized to  her body or location, because she is thereby measuring a \emph{complete} relational  variable pertaining to the relation between her and the entity she is measuring. Therefore what she is measuring is in fact physically real even though partial observables on their own are not physically real - partial observables are simply how complete observables look `from the inside,' so to speak. 

For example, \cite{rovelli2022philosophical} write, `\emph{The quantity measured by the detector is a local function of the metric, in the location determined by the detector. The full diffeomorphism invariance of the pure gravity dynamics, in other words, is physically broken by the detector itself being located somewhere.}' But of course, measurements are ultimately performed by observers, not detectors, so the breaking of the diffeomorphism invariance by the position of the detector does not fully solve the problem: in order to observe a reading on the detector, the observer must presumably do something like inspect the position of a pointer on a dial, and the position of a pointer is again a partial observable,  so it seems the observer should not be able to observe it. Therefore the mere existence of some relation between a detector and a local function of the metric is not enough to explain the possibility of observing relational observables: somehow the observer must actually get the information about that relation into her conscious awareness.  This line of reasoning suggests that perhaps the description offered by Rovelli and Vidotto might ultimately be applied to the observer -  the diffeomorphism invariance could finally be broken by the \emph{observer herself} being located somewhere. After all, this does seem to correspond to how we actually make observations: when I notice the positions of objects around me I am not localizing them relative to some abstract coordinate space, or even relative to a measuring device, rather I am noticing how close these objects are to me, and how they are oriented relative to my standpoint. That is, I am observing their position relative to my own position - and that relation is, of course, a total observable rather than a partial one.

\subsection{Partial Observables Relativized to Oneself \label{oneself}}

\cite{Callender2017-CALWMT}, echoing a common sentiment in the philosophical study of time, writes that in studying the relation between our usual understanding of time and true physical time, `\emph{I shall not try to close the gap between temporal \textbf{experience} and physical time. That would be to try (in part) to solve the mind-body problem, and my aspirations are far less grand ... Deducing temporal experience from physics is a fool's errand.}' But if the relational observables approach is right, it is not clear that we can study physical time without telling  some story about consciousness and temporal experience. For the relational approach suggests that physical time itself is in some sense a product of our internal perspective - the variables involved in defining physical time have no independent reality, and thus they can only be defined from the inside, within our view on some complete relational observable.  If this is correct, we are not going to be able to tell any story about physical time, or indeed any other aspect of empirically observable physics, without saying something about the nature of the internal perspective relative to which variables are defined.

That said, as noted in section \ref{intro}, we do not  need a complete solution to the mind-body problem for this purpose - rather what we need is  a schematization of the observer, i.e. we need to identify at least in approximate terms what the physical correlates of consciousness could possibly be in a diffeomorphism-invariant theory. Curiel emphasizes that a meaningful schematization may be quite simple and abstract, and in accordance with this I will not appeal to specific details of neuroscience or our perceptual apparatus in this paper;   I will simply try to identify a very general, high-level class of variables which could plausibly play the necessary role. 

To begin with, I will  adopt  a necessary condition for the content of conscious experience, which   I refer to as the  `Reality Criterion':  \emph{conscious experience cannot supervene on or have direct access to any fact about the world which is not physically real}. Note that this criterion is written in such a way as to be neutral between physicalist and dualist accounts of consciousness - physicalists would  imagine consciousness as supervening on some collection of physical facts, while dualists would regard consciousness as separate from physical reality but as having direct access to some particular collection of physical facts. Thus the Reality Criterion seems like a requirement that would be considered broadly reasonable across many different views on consciousness. 

 Now, if  we adopt the standard interpretation of a gauge theory, a corollary of the Reality Criterion is that consciousness cannot supervene on or have access to any fact which is not a (complete) observable. One might perhaps be inclined to object to the latter claim on the basis that what is `observable' is a function of the physical limitations of our sensory system, whereas  consciousness presumably does not supervene or have access to physical reality by means of sensory perception, so the relation between consciousness and reality should not be constrained by observability.  However,  that would be to confuse the technical term `(complete) observable' in the gauge theory context with the usage of this term in everyday language - it is true that what is observable in the everyday sense is a function of the perceptual capacities of the observer, but the standard interpretation says that what is  `observable' in the gauge theory sense includes everything that is physically real, so within the standard interpretation  the Reality Criterion does indeed imply that consciousness can only supervene on  (complete) observables. 

So let us  see what the Reality Criterion  has to say about the process of observing a partial observable relativized to one's own state or location.  I will consider two possible approaches: an externalist approach in which we say that consciousness supervenes  on or  accesses the  total observable associated with the between the brain and the external \emph{environment}, and an internalist approach in which consciousness can only supervene on or access total observables which are internal to the brain. We shall see that the externalist approach seems difficult to reconcile with ordinary intuitions about conscious experiences, while the internalist approach seems more intuitive, but requires us to do more work to identify the relevant kind of total observables.

 \paragraph{The Externalist Approach}

Roughly speaking, we usually imagine that what happens during an observation is that   the brain of the observer becomes correlated in some way with some feature of the environment. And   a  correlation between the brain and some feature of the environment is clearly a  `relational' fact, so it seems likely that such a correlation could   be expressed as a complete relational observable, meaning that the Reality Criterion would allow consciousness to supervene on or have access to such relations.  This would be a straightforward way in which we could come to observe relational observables: our consciousness could supervene on or have access to total observables in the form of relations between our brains and features of our external environment. 

However, it should be noted that if we say that consciousness supervenes on or has  direct access to the relation between the brain and an external object,  we are then compelled to adopt an externalist account of consciousness, i.e. a view in which conscious experience may depend on something outside of the body. Indeed, relational approaches to physics are often presented in a way which seems to presuppose some kind of externalism - for example, people working in the QRF formalism commonly refer to an description of the external physical world relative to a given observer as the `perspective' of that observer (see for example  \cite{2020acop}, which could potentially be read as endorsing a version of externalism). 

Now, there are  good reasons  to believe in \emph{some} kind of externalism about mental phenomena. In particular, it is commonly argued that the `content' of an experience - i.e. its meaning or reference, or what  it represents about the world - must depend on external features of the individual's environment and history,  in addition to intrinsic features of the individual such as the state of their brain \citep{Putnam1975-PUTTMO,Burge1979-BURIAT-11}. 

However, what is at stake in the question of how we come to observe relational observables is not `content' in this intentional sense. For one major reason why it is important to understand how we observe relational observables is because  a satisfactory epistemology of science requires a reasonable grasp of the way in which we obtain the information that we use in our scientific reasoning. So we are interested not in the intentional content of experience, but merely in  the availability of physical information, in a thin sense - a variable is available to some observer in this sense  if it is possible for that observer to take a (consciously chosen) action or to engage in (conscious) reasoning in a way which depends on the value of that variable.  And in order for a variable to be available to an observer in this pragmatic sense, the value of that variable must in some way be instantiated in their `access consciousness,' which is a term used by Block to refer to the part of  phenomenal experience which   is `\emph{(available) for use in reasoning and rationally guiding speech and action}' \citep{Block1995-BLOOAC}. So in order to argue that we  observe relational variables in virtue of our consciousness supervening on external relations between the brain and the environment, it would be necessary to adopt   externalism not only about intentional content, but also about access consciousness - that is, it would be necessary to say that observers can sometimes access and consciously act on information which is not present in any part of the complete history of their brain.

Externalism about access consciousness   would have  a number of odd consequences. First, such an approach seems hard to square with a modern scientific understanding of perception. We have  detailed knowledge of the functioning of our sensory organs, which reveals that they  go to considerable trouble to get information about the external world `into' the head so that we can use it in reasoning and guiding action; why would they need to do that if the accessible part of our phenomenal experience is able to depend directly on external facts? Additionally, modern neuroscience reveals that the accessible part of phenomenal consciousness is  closely correlated with internal states of the brain -  \cite{Pautz2019-PAUWIT-2} offers a detailed discussion of this evidence, noting that phenomena like pain intensity, smell and taste, audition and the perception of colour all seem to be related in a direct way to corresponding  states of the brain. While it remains possible that there is some subtle external dependence that is missed by these experiments, nonetheless the evidence seems to point strongly towards much of phenomenal experience - and certainly the \emph{accessible} part, which is after all the only part that can be reported in these experiments -  being determined by factors internal to the brain.

 Externalism  about access consciousness would also potentially lead to problems regarding locality. Suppose it is true that when I observe an apple, the external relation between the apple and my brain directly determines   the accessible part of my phenomenal experience. Then  if the apple spontaneously combusts, it seems that the accessible part of my phenomenal experience should instantaneously change in some way, even though locality suggests it should take some time for the information about this event to reach me. Moreover, by definition access consciousness can be used in guiding action, so if we assume that I can act on it more or less instantaneously, it seems as though I should be able to react to this combustion at a point outside of the future lightcone of the combustion event, which would look like superluminal signalling. And of course, to describe this process physically we would have to pick a preferred reference frame to define the moment at which the accessible part of my phenomenal experience changes and my body performs the corresponding action, in violation of the relativistic idea that there are no preferred reference frames. So certainly the simplest version of the externalist approach seems quite incompatible with relativity.
 
 In summary, externalism about access consciousness does have a certain appeal  within a relational approach to physics, and perhaps this approach merits further investigation, but at present there appear to be a number of obstacles. Thus in this paper I will focus on trying to understand how to reconcile the relational approach with an internalist approach to access consciousness.
  
\paragraph{The Internalist Approach}

An internalist view of access consciousness would involve postualting that the accessible part of our phenomenal experience  supervenes on, or has direct access to, only internal facts about  the brain. This internalist view of  access consciousness seems more consistent  with our modern understanding of the relation between brains and conscious experience. However, it  leads to a difficulty for relational approaches. For  the idea that I can directly observe a partial observable relativized to my own state or location seems to assume that what I am observing  is outside my body and distinct from me, so that my consciousness then supervenes on or accesses a complete relational observable, i.e. the relation between my  body and the external system.  But if the information about this partial observable has to be copied into my brain before I can become conscious of it, the relational aspect seems to have been lost - the variable of which I am aware is no longer a relation between my body and an external object, it has just become a non-relational variable characterising the state of my brain in some spacetime region, and thus the Reality Criterion suggests my consciousness should not be able to supervene on or access it. 

For example, suppose I am trying to measure the distance between myself and my cat. That distance is a complete relational variable, but normally we would imagine that in order to actually measure it I have to use a measuring instrument and then observe the instrument, which presumably amounts to copying the reading from the instrument onto something like a `register' in my brain. But the state of the register in my brain which records the distance does not look like a complete relational observable any more: it is just the state of a physical system at a time, so it looks rather like a local function of coordinates. Such a thing cannot be a complete observable, and therefore the Reality Criterion suggests that consciousness should not be able to supervene on or have access to it. 
  
Now, one possible solution here would be to resist the intuitive pull of diffeomorphism invariance and argue that consciousness \emph{does} in fact have some kind of transcendental access to spacetime regions and their contents. For example,  perhaps a consciousness is somehow  attached to a spacetime region, and what occurs in its conscious awareness is a function of the state of the brain in that region. This would amount to denying the connection between `predictable' and `physically real' in the standard interpretation of gauge theories - in this picture, the values of variables at individual spacetime points \emph{are} physically real because consciousness can supervene on them, even though they could never be predicted by a diffeomorphism-invariant theory. However, this approach seems hard to accept - it would seem to entail that there are consciousnesses in arbitrary regions regardless of the presence or absence of a brain there, just sitting waiting for something to attach to. It seems much more natural to think of consciousness as being attached to brains or other suitable physical structures,   and thus it is natural  to expect that   our conscious experience - over the full span of our lives, rather than at a pre-specified time - should be invariant under diffeomorphisms. 
 
This suggests that in order to make the internalist view work, we will probably need a model which ensures that when a partial observable is observed, this is achieved by converting it into a complete observable defined entirely within the confines of the brain, which consciousness  supervenes on or has direct access to. Moreover, as noted in section \ref{intro}, complete observables fall into two classes - highly non-local complete observables, or complete relational observables - and since highly non-local observables cannot  be defined wholly within the confines of the brain, it seems likely that the relevant kind of observables are, once again, relational ones. So  the task before us is to understand  how facts about the external world could be converted into  complete relational variables existing entirely within the confines of the brain, in order that consciousness could have knowledge of or supervene on them.

\section{Agency \label{action}}

The QRF formalism  and the Page Wootters approach suggests an obvious way to identify an appropriate set of complete relational variables within the confines of the brain. For we know that  in order to recover  ordinary time evolution in the reference frame formalism we need to use something like a clock as a reference frame, and thus it is tempting to   associate with each brain some kind of internal clock, so we can  formulate complete relational variables of the form `this brain is in state $\psi$ conditional on its internal clock being in the state $t$.' I will refer to variables of this kind as `$\psi$-at-$t$ variables.' So one might imagine that consciousness has direct access to or supervenes on $\psi$-at-$t$ variables, defined entirely within the confines of the brain, meaning that we can be aware of something like the temporal evolution of our brain\footnote{  \citep{Fields_2022} discusses an alternative way of using the QRF formalism within the brain, by understanding neurons as hierarchies of quantum reference frames. However, this work employs  the quantum information approach \citep{RevModPhys.79.555}, rather than the QRF formalism as discussed here.  In addition, my interest here is in identifying some kind of variable which consciousness could supervene on or have access to; and we would not want to say that consciousness supervenes on or has access to relative variables defined relative to all of the neurons in the brain \emph{individually}, because this would either lead to too many consciousnesses or to a highly disunified consciousness which does not respect our actual experience - somehow such relations must be unified into a higher-level relational variable before we become conscious of them.}.

 However, there is  still a  puzzle. For   as \cite{Earman2002-EARTMM} emphasizes, `coincidence variables' or complete relational variables are timeless   - if it is true that `this brain is in state $\psi$ conditional on its internal clock being in the state $t$' then this is \emph{always} true, timelessly and eternally. So if consciousness supervenes on timeless variables like this, it seems quite unclear how we come to have the experience of being located \emph{in} time. After all, there is no obvious reason why awareness of one $\psi$-at-$t$ variable should be mutually exclusive with awareness of other $\psi$-at-$t$ variables -   surely I could simultaneously be aware that `this  brain is in state $U_1$ conditional on its internal clock being in the state $t_1$' and also  that `the brain is in state $U_2$ conditional on its internal clock being in the state $t_2$'. So if our experience supervenes on these timeless relational variables, why is it that we only ever experience one moment at a time, rather than  experiencing some kind of `all-at-once' knowledge of our whole life's timeline?

This issue is related to the problem of time in quantum gravity, and indeed it has been suggested that problems of this kind undermine the claim of the partial observables approach to have fully solved the problem of time. For as \cite{pittphilsci15795} puts it, `\emph{(if)  the partial observables are non-measurable, then we seem to lose our ability to use different values of the internal clocks to describe change. Rather, all we have are measurements of the complete observables which are (in a precise sense) temporally non-local.}' That is, the possibility of experiencing oneself as localised at some specific time is clearly a prerequisite for experiencing anything which could be described as change - if you always perceive the whole of history at once, you cannot possibly experience change. And yet if reality is composed of nothing but timeless complete relational variables, like the $\psi$-at-$t$ variables, it is not obvious how any agents could possibly experience themselves as being located at a specific time. So in order for the relational observables program to fully resolve the problem of time, it must be able to provide an answer to this question about how  observers experience themselves as being located in time.  
 
Here is helpful to consider a thought experiment exploring what it \emph{would} be like to have `all-at-once' knowledge of several different moments in your life's timeline, rather than experiencing yourself as localised at a time - thought experiments of this kind have  been explored in fiction \citep{chiang2010stories,vonnegut1998sirens} and theology \citep{boethius2016consolation}. One obvious consequence of having knowledge of several different times is that it would necessarily place limitations on agency. For as \cite{Callender2017-CALWMT} puts it, `\emph{Part of what it is to be an agent is to have this sense of freedom, a sense that other future options are in some sense live.}' And if you already know what you are doing at time $t_2$, that significantly constrains the extent to which future options can still be live at an earlier time $t_1$. For example, if I know that on Tuesday I will be eating a cake I baked on Monday, I cannot possibly refrain from baking a cake on Monday, since I will not be able to eat the cake on Tuesday if I did not bake it. So there is necessarily some kind of tradeoff between knowledge and agency: the more detailed your knowledge of what is going on at $t_2$, the more limited your range of possible actions at $t_1$.

Therefore a being with `all-at-once' knowledge of its whole life's timeline would necessarily be very different from us with regard to its relation with causation and agency - as \cite{McKenna2023-MCKAAT-10} notes, `\emph{were an `eternalist perceiver' ... to exist in our universe, this entity would be hamstrung by a temporal perspective that entails a degree of causal and epistemic impotence.}' Similarly, \cite{doi:10.1080/24740500.2022.2155200} observes that a tradeoff between knowledge and agency is present any time a reasoning system like a computer is asked to predict its own future actions:  `\emph{By saying things, the computer is doing things and that keeps it from being able to stabilize the facts it is representing independently of how it represents them.}' She introduces a concept of  `negative interference'  to describe this tradeoff, referring to the fact that an agent cannot ever have stable knowledge of their own  future actions. As \cite{doi:10.1080/24740500.2022.2155200} emphasizes, any physical system engaging in any kind of self-reference is necessarily subject to this negative interference: `\emph{Any system that is acting in the domain it is representing is going to encounter interference ... Interference impedes pure knowledge acquisition.}' 

It is a corollary of Ismael's account of negative interference that if an agent is  contemplating an action at time $t$, that agent cannot in general have reliable knowledge of their own states at times later than $t$, since such knowledge will typically imply something about the action taken at time $t$, and therefore  negative interference ensures that the agent cannot have such knowledge. For example, in the case described above,   if I am still deliberating about whether to bake a cake on Monday, then I cannot possibly know that  on Tuesday I will be eating the cake I baked on Monday, because if I knew that and then chose to refrain from baking a cake on Monday this would lead to a physically  impossible sequence of events.   

A further corollary is that agents cannot simultaneously make choices about actions which occur at two separate times $t_1$ and $t_2$ on the same worldline, because certain choices of action at $t_2$ will imply something about the action taken at time $t_1$, and negative interference ensures that the agent cannot have such knowledge. For example, in the case described above, if I am  simultaneously deliberating over my actions on Monday and Tuesday, then I should be able to decide   both to eat the cake on Tuesday and to refrain from baking it on Monday, leading to a physically impossible sequence of events. As McKenna puts it, for an agent with epistemically symmetrical knowledge of two times on the same worldline, `\emph{no process of deliberation takes place, no prediction, no decision making, no attempt to bring about an event, and no selection process. These actions all presuppose the utilization of past information in service of enacting future outcomes in an unrealized future. Actions of this type are not only rendered irrelevant by epistemic symmetry but would be impossible except as charade}' \citep{McKenna2023-MCKAAT-10}. Thus it seems that genuine agency in the sense in which we usually understand it does need to be localised at a specific time or in the performance of a single deliberation, and therefore taking account of the relation between experience and agency starts to give a  clearer sense of why agents must necessarily experience themselves as being located `in' time.

\subsection{Passive awareness vs Agency \label{agency}}

The philosophical literature on temporal experience often seems to presuppose something like a `passive awareness' conception of consciousness, in which our experience has an epiphenomenal character. For example, one common account of  the temporal asymmetry of agency suggests that it is merely an emergent feature of reality  which follows from the thermodynamic gradient as encoded in the second law of thermodynamics \citep{Huckleberry,Albert2000-ALBTAC,kutach2013causation}. Specifically, this idea is often cashed out in terms of an epistemic asymmetry:  if we assume that entropy was low in the past, as stipulated by the second law, it follows that the current states of environmental systems will carry traces which give specific detailed information about events in the past, but they will not typically carry traces giving similarly detailed information about the future. Thus an observer at $t$ who has information about these current states will necessarily experience the past relative to $t$ as fixed but the future relative to $t$ as open, simply because there is more information available to her about the past than about the future. This is most naturally understood in terms of   a `passive awareness' account of conscious experience, since it essentially reduces the availability of action to awareness or observation. 

Yet there is something oddly one-sided about the passive awareness account, for it assumes that \emph{awareness} is real, but \emph{agency} is not. That is, our feeling that we are aware of the states of things in the world is veridical, but our feeling that we act on the world is some kind of illusion which results entirely from the structure of the things our awareness happens to supervene on - i.e. brains with memories of the past and not the future. Yet it is not obvious that awareness is more fundamental than agency from a phenomenological standpoint. 
Indeed, \cite{Young2022AgentsOC} argues that `\emph{the sense of ourselves as causing change is an experiential constant. We feel ourselves as effecting change at every instant we are conscious,}' noting that `\emph{the moments in which we are not purposefully controlling our bodies are less common than we might think. Even as you sit reading these words you are still purposefully keeping your body rigid enough for it not to slide out of your chair, and are moving your eyes in such a way as to facilitate reading, rather than letting them glaze over or drift around the page.}' \cite{Young2022AgentsOC} also points out that  `\emph{the body could be entirely at rest and yet we can still deliberately deliberate on a problem, purposefully bring a memory to mind, or privately curse in inner speech.}'  The close connection between consciousness and action has also been emphasized through the embodied cognition research program\cite{sep-embodied-cognition}. So perhaps what is really illusory here is not our experience of agency, but the common idea that we can  break awareness and agency apart. 

Standard treatments of the mind-body problem may have contributed to this idea. For in this context, much effort has been expended on arguing that there is something more to consciousness than mere behaviour - for example, the famous zombie argument contends that there could exist `zombies' who behave just like human beings but without conscious awareness \citep{sep-zombies}. And this emphasis on the fact that action does not fully explain awareness leads naturally to the impression that action is completely distinct from awareness. However, even if we accept the zombie argument and other related arguments, it does not follow from the fact that it is in principle possible to have behaviour without awareness that it is also possible to have awareness without agency, or at least the possibility of agency.  Moreover, even if  a non-agential consciousness \emph{is} possible at least conceptually, such a thing would clearly experience time in a very different way to us, so in seeking to solve the experiential problem it makes sense to employ an understanding of consciousness which gives agency a prominent role. 

So perhaps some of the obstacles that have arisen in attempting to reconciling the Reality Criterion with the nature of our temporal experience may arise from an overly narrow focus on consciousness as passive awareness. We have been trying to account for temporal experience by identifying some set of diffeomorphism-invariant variables  which consciousness could supervene on or or have direct access to, such as the $\psi$-at-$t$ variables, with the implicit understanding that this amounts to some kind of epiphenomenal passive awareness of these variables. But we are not just   passive observers; we are also agents. And the experience of being located at a time is not just the experience of being aware of the state of the brain at that time, but also of having available to us a variety of forward-facing actions.  As we saw in section \ref{action}, taking this feature of conscious experience into account may help explain why  agents experience themselves as being localised \emph{at} times, and indeed, it's  likely that an agential conception of consciousness  could help make sense of other aspects of our temporal experience as well - for example,  \cite{doi:10.1080/24740500.2022.2155200} sees agency as part of the explanation for why we experience the future as open, and \cite{Young2022AgentsOC} suggests that `agentive experience' is an important part of the reason why people commonly believe that time passes or flows. 
 
Now, one may naturally worry that in the context of modern theories of physics like special and general relativity and quantum gravity, which are arguably most compatible with a block universe picture\cite{10.1093/oso/9780198807087.003.0003,Putnam}, there is   not really conceptual space for consciousness to involve anything more than awareness of the values of various local functions of coordinates. After all, how can agents  be actively involved in producing variables which already exist atemporally and eternally?  However, if the proponents of relational solutions to the problem of time are correct, we are in fact being pushed by  modern physics towards an account of physical reality which is still a `block universe,' but  one whose empirically accessible contents are \emph{relations} rather than local functions of coordinates. And as we will see in the next section, the relational view seems more friendly to an agential model of consciousness than a comparable block universe view based on local functions of coordinates, so the relational approach may offer a way of reconciling the block universe picture with an agential conception of agency, thus demonstrating   that the block universe  need not be synonymous with a passive awareness conception of consciousness.

\subsection{Relational Agency}

Suppose  we accept the idea that a satisfactory account of temporal experience requires an agential conception of consciousness. What does that imply for the problem of identifying relational variables which consciousness could supervene on or access? Well, the initial idea that   consciousness  supervenes on or has direct access to something like   $\psi$-at-$t$  variables seems to be rooted in the passive awareness conception of consciousness, in which it is often supposed that consciousness supervenes on or has knowledge of  the state of a brain at a single time. Whereas if we move to an agential model, there may be other kinds of relations which are a better fit for the physical origins of consciousness. 

For example, the `event-causal' theory of agency \citep{60a9dd9a-e48a-3fcf-87dd-d5c158406fc6} suggests that an intentional act must involve a certain kind of causal relation between mental states like desires and beliefs, and actions. So if we are working in a relational approach, in which we have already accepted that consciousness must supervene on some kind of relation, we have a natural route to an agential conception of  consciousness by simply suggesting that consciousness   supervenes on or perhaps simply inheres in this kind of causal relation.  Relations of this kind look  more promising than $\psi$-at-$t$ variables as a way of   accounting for our actual temporal experience, which is not of passively noting states relative to clocks, but rather of receiving information and immediately acting on it\footnote{Obviously, there is still a question about which specific relations of this kind are associated with consciousness - we presumably do not wish to say that all causal or quasi-causal relations are linked with consciousness. One might imagine that this could be done using some criterion which identifies relations with a certain kind of complexity, perhaps by appealing to some theory of consciousness such as integrated information theory\cite{tononi}, but I will not try to address this question here.}.  Of course, it seems likely that such relations must be understood as temporally extended or modal in character - indeed, perhaps both - but this is not necessarily an obstacle, since it has often been noted that the experiential present seems in some sense to have a finite duration \citep{Andersen2014-ANDTDO-3}. 

 So suppose we say the physical origins of consciousness are found in complete relational observables in the brain, which relate  inputs in the form of memories and sensory impressions to outputs in the form of  decisions. The inputs and outputs are partial observables, so they are not individually physically real: it is the relation between them which is a complete observable, so it is only this relation which is to be invested with physicality. Now, at this juncture one might object that just as consciousness surely cannot supervene on variables which   are not physically real, a process of deliberation surely cannot take inputs which are not physically real\footnote{Thanks to an anonymous reviewer for raising this point}. But this objection is predicated  on a picture in which awareness is separate from action, such that in the process of deliberation we \emph{first} become aware of the inputs and then \emph{subsequently} engage in some deliberation on that basis. Yet as argued in section  \ref{agency}, it is not clear that  awareness and action can really be taken apart in this way. And if they cannot be cleanly separated, then we should not think of a process of deliberation as depending on  an input which has  physical reality separate from the process itself:  rather the input, the output and the relation between them must be understood holistically, because the input and output are not individually physically real and therefore the process cannot be further decomposed into individual elements. This holistic approach to deliberation ensures that the whole process can be described in terms of total observables, rather than relying on partial observables.

Now, of course we normally think of deliberation as having some internal direction which determines the difference between inputs and outputs, so in the picture I have just suggested there is a natural question about the origins of this direction of deliberation.  In accordance with the modern scholarship on the arrow of time, I think it is likely that this time asymmetry has something to do with   thermodynamics: the question of how to properly represent thermodynamic effects in a theory based on relational observables is clearly interesting and important, but this issue is beyond the scope of the current paper, so for now I will simply assume that the temporal asymmetry is indeed inherited from thermodynamic effects. Note that as discussed in \cite{adlam_2023}, deliberative processes must of course be linked together in consistent ways, and there is some reason to think that the requirement of consistent chaining is closely tied to the causal structure of spacetime, so investigating the connection between deliberative relations and agential consciousness  would be an interesting area for investigation in future work. 

Using the idea that consciousness supervenes on or accesses complete relational observables in the brain which represent processes of deliberation, we can construct a story about how it is that we observe relational observables. That is, an observer Alice can observe a partial observable in the environment, relativized to her physical body, by a process that involves several steps. First Alice's brain becomes correlated with a partial observable $P_1$ in the environment, giving rise to a complete relational observable relating $P_1$ to a corresponding partial observable $P_2$ in Alice's brain. For example $P_1$ might be a reading $t$ on a clock $C$ at some spacetime point $x_1$, and $P_2$ would be a record of that reading at some spacetime point $x_2$ inside Alice's brain. Then there is a process of deliberation in   Alice's brain which takes as input $P_2$ and produces as output a decision  that she will measure some variable $V$ of a system $S$. That decision then produces external effects - bodily motions and so on, which act on the system $S$ at some spacetime point $x_3$ and produce a measurement outcome which is a partial observable $P_3$ encoding information about the value $b$ of the variable $V$ at $x_3$. Finally sensory perception is used to correlate Alice's brain with the external reading on the measurement device, giving rise to a fourth partial observable $P_4$ in Alice's brain which is a record of the outcome $v$ at some spacetime point $x_4$ inside Alice's brain.  Thus if we restrict attention to matters internal to the brain,  we can identify something like two chained complete relational observables: the relation between the partial observable $P_2$ and the later decision to act, and the relation between the decision to act and the later partial observable  $P_4$. So  if we think of Alice's consciousness as supervening on or having direct access to these internal complete relational variables, then what Alice will observe is something like `the value of $V$ is $v$ relative to the clock $C$ showing time $t$'. Thus the specific spacetime locations of these variables are not relevant to Alice's conscious  experience, since the variables  are  bound together in her Alice's awareness simply by her personal involvement in the measurement process, and therefore the facts she is actually aware of are  diffeomorphism invariant. Thus this picture ensures that   consciousness ultimately inheres only in complete relational observables, so it obeys the Reality criterion.

 Indeed, in a sense this agential story about how we observe partial observables is already implicit in their original definition. For   \cite{Rovelli_2002}  describes them as variables `\emph{to which we can associate a measuring procedure leading to a number}'- so Rovelli's conception of partial observables is not that we are passively aware of them, but that we \emph{act} on them, carrying out a measuring procedure which leads to a number. And an idea  of this kind also seems to be implicit in Maudlin's operational description of how we measure tensor fields in spacetime. \cite{Maudlin2002-MAUTMM} writes that in order to measure the Ricci tensor at the point where the rockets collide, one would not measure the Ricci tensor and then separately check for collisions of rockets and then verify that they coincide: rather `\emph{one would tell by sending a rocket along each path and making the measurement when they collide.}' This is very similar to the procedure I have just described: we simply replace $P_1$ with the fact that the rockets crossed at some spacetime point $x_1$, and we replace $P_3$ with the reading of the device measuring the Ricci tensor at $x_3$, so the observer involved makes an observation of the form `the value of the Ricci tensor is $r$ relative to the rockets colliding,' and this description corresponds to a  manifestly diffeomorphism-invariant complete relational observable internal to the brain. Therefore modelling consciousness in this agential way may offer a clearer path to making sense of the actual physical process we go through when we observe relational observables - we do it not by passively looking in at some partial observable, but by literally being part of the complete relational observables involved in the process of observation.

\section{Quantum Relational Observables\label{quantum}}

 Having examined the classical case, let us now try to understand how this analysis may change if we are interested in a diffeomorphism-invariant theory of quantum gravity rather than classical GR. We should begin by considering the nature of the complete relational observables which consciousness must supervene on or have access to in this picture. For example, suppose we try to formalise these variables using the QRF formalism, in which   we can simply switch into the reference frame of a given system observer and then  define  variables of the form `$X$ is in state $\phi$ relative to $Y$ being in the reference state $e$'. Since these variables are stated relationally, it is reasonable to expect that they correspond to total observables and thus the Reality Criterion would allow consciousness to supervene on or access them. 
 
 However,  we should examine more closely the phrase `relative to.' Phrases of this kind are used very frequently in the QRF formalism (see also `conditional on,' `given that' and `in the perspective of' and so on). But as noted in   \cite{AdlamPW}, their meaning is somewhat opaque: in ordinary language these phrases often mean something like `at the same time as,' but that  can't be what is meant in this case. After all, if consciousness has to become aware of these relational variables by first becoming aware that  $X$ is in state  $\phi$ in a spacetime region with time coordinate $t'$, and then becoming aware $Y$ is in state $e$  in some other spacetime region with time coordinate $t'$, and then noting that these things have the same time coordinate $t'$, it looks like we are after all violating the Reality Criterion:  `the state of the brain in a given spacetime region' and `the state of the clock in a given  spacetime region' are not observables, so consciousness presumably cannot be aware of them separately. 

So let us look more closely at how these relational variables are constructed. This is done by first taking some state $\psi$ of the joint system of $S$ and $T$  as a whole and then applying a symmetry transformation to this state, leaving us with a physically equivalent state which can be written in the product state $| \phi \rangle_X \otimes | e \rangle_Y$, where $| e \rangle_Y$ is the unit element of the symmetry group; we then say that $\phi$ is the state of $X$ relative to $Y$ being in the state $e$. So what links $\phi$ and $e$ together seems to be simply the state $\psi$, or rather some object corresponding to the equivalence class to which $\psi$ belongs when we quotient the set of possible states by symmetry transformations. This object is `physically real' in the sense of being invariant under gauge transformations, but it cannot be cashed out as representing any particular spatiotemporal relation, since the original state $\psi$ need not involve  any particular spatiotemporal relation.  Indeed, \cite{Lam_2023}. , examining specific instances of relational observables appearing in  loop quantum gravity and causal set theory,   suggest that `\emph{these connecting threads (must) be understood as non-spatial relations of joint co-existence in the same structure.}'
 
Moreover,   \cite{Lam_2023} argue that it seems implausible that these `threads' can be understood as representing some kind of non-modal relation. For if they were non-modal, we would naturally expect that they could feature in something like a Humean supervenience basis, which means they should obey a principle of free recombination - that is,  we should be able to rearrange the elements of the supervenience basis in any way  we like and still arrive at a  metaphysically possible world. But  \cite{Lam_2023} argue that  the structures encoding relational observables probably cannot obey anything like a principle of free recombination, because  this structure can only be meaningfully defined within a constrained Hilbert space formalism, so it  doesn't make sense to try to recombine these relations in ways forbidden by the kinematics of the underlying Hilbert space. That is, the kinematics appears to encode `necessary connections' which must exist prior to any systematization of the supervenience basis, since otherwise we could not even define the supervenience basis in the first place. So in the quantum context, if we accept that the substantive physical content of the theory is in the complete relational variables, we appear to be pushed towards a picture of reality which in some ways looks almost exactly opposite of the Humean picture: reality appears to be composed of a network of  modal or nomic relations, with localized non-modal properties  appearing as `partial observables' which are not individually real, but which simply emerge from a particular perspective on this network of modal relations\footnote{As noted by  \cite{pittphilsci4223} - though in the context of GR, rather than quantum gravity -  this could potentially be interpreted as a form of ontic structural realism.}. 

And if it is true that in the context of a quantum theory of gravity we need to think of consciousness as supervening on fundamentally modal relations rather than purely spatiotemporal ones, that seems favourable for the agential conception of consciousness. For agency is, in essence, a modal notion: a process of deliberation does not constitute meaningful agency unless there is some genuine dependence of the output on the input, mediated by specific features of the agential process. In particular, I argued earlier that in an agential conception of consciousness we might think of consciousness as supervening on something like a causal relation in the brain, and causal or quasi-causal  relations are usually understood as being modal in character. So  if the nature of relational observables in the quantum context  is in any case pushing us towards a non-Humean picture based on fundamental modal or nomic relations,  that means we have the resources to give agency its due - rather than being a mere illusion parasitic on an epistemic asymmetry, it can inhere in a modal  structure which is  physically real according to the theory and indeed constitutes the main substantive physical content of the theory.

\section{Quantum Agency}

In section \ref{agency} I argued that it may be easier to explain our temporal experiences and in particular our experiences of being localized at times if we adopt an agential conception of consicousness rather than a passive awareness of consciousness. This argument seems to apply equally well to the quantum case as to the classical case, and therefore there is some reason to think that giving a full solution to the experiential problem in the quantum context will require us to employ some quantum account of observation, deliberation and action. Moreover, the quantum problem of time is usually formulated in a context where it is assumed that quantum mechanics   is universal, so if we hope to use our solution to the experiential problem to address the problem of time,  we can't retreat to a separate classical domain to describe the process of observation or action - we will have to be able to model these processes in entirely quantum terms. 

It should be emphasized that simply appealing to the usual operational interpretation of the quantum state does not  solve this problem. One might be tempted to suggest that in the relational context, an agent  simply becomes aware of  some value for a variable with a probability proportional to the corresponding mod-squared amplitude in the quantum state of the relevant system relative to her  reference frame. However, this will not work if we accept that an  agential model of consciousness is needed  - for in that case we must be  able to go on modelling the agent \emph{after} her observation is complete, in order that  we can describe her as engaging in a process of deliberation which takes the result of her observation as an input. So we can't just assign probabilities to measurement outcomes;  we also have to specify a post-measurement state for the observer and the measured system, and this necessarily requires us to go beyond the operational interpretation to a substantive physical account of the process of making and acting on an observation.

It is  straightforward to give such an account from an external point of view. For example, suppose   Alice performs a measurement in the basis $\{ |X \rangle \langle X|, |Y \rangle \langle Y | \}$ on a quantum system $S$. The standard quantum description, applied from the  point of view of some external observer,  tells us that Alice and the system will now end up in a joint superposition state, which might look something like $\psi = \frac{1}{\sqrt{2}} | O_X \rangle_A | X \rangle_S + \frac{1}{\sqrt{2}} | O_X \rangle_A | Y \rangle_S$, where $|O_X\rangle$ is the state corresponding to Alice seeing outcome $X$ and $|O_Y\rangle$ is the state corresponding to Alice seeing outcome $Y$. But  if we trace over system $S$, Alice's state  is just $Tr_S( \psi)$, which is a mixture of the $X$ outcome and the $Y$ outcome, so from the external point of view it seems as though she has not seen any definite outcome to her observation at all. And we are not going to be able to locate Alice's conscious experience of time in her internal process of deliberation if we cannot first obtain a definite outcome for her to deliberate on, so we need some way of extracting such a thing out of the formalism. 

This, of course, is essentially just a version of the measurement problem. Now, one might perhaps  hope to sidestep the measurement problem by making use of the QRF formalism - that is, one might hope that we can simply shift into Alice's QRF and then model her experiences and her process of deliberation and agency within that internal reference frame. Initially this approach  looks promising - by definition   Alice cannot be superposed in any basis relative to her own reference frame, so it seems that when we move to Alice's reference frame we end up with a description in which Alice is having a single definite experience, even though relative to Bob she appears superposed.  Moreover,  people working in the reference frame formalism often use the term `perspective' to refer to reference frames, and describe the process of switching into a reference frame as `jumping into the perspective’ of a quantum system, which naturally  gives the impression that in the case of a conscious observer, the content of the reference frame is supposed to have something to do with what the observer is experiencing. 
 
However, unfortunately using the QRF formalism presents difficulties insofar as our goal is to model experience based on an agential conception of consciousness. For the motivating idea of the formalism is that the relevant degrees of freedom of the reference frame itself are removed from the physical description relative to that reference frame. For example, in the case where we are dealing with position and momentum references QRFs, then `\emph{the position and momentum of the reference frame are not dynamical variables when considered from the reference frame itself ... the reference frame is not a degree of freedom in its own description, but external systems to it are}’ \citep{article3}. And removing Alice's position and momentum as dynamical variables seems to be effectively equivalent to removing Alice from our description altogether. After all, if we cannot model Alice as having a position which changes in this reference frame, it would seem that we will not be able to model her as taking action, since effectively all forms of human action involve some kind of motion. So there seems to be no possible way of describing an observer as an agent using their own reference frame as defined within the current QRF formalism, since the degrees of freedom of the observer which are relevant to the description are always removed from the relativized description.  

Thus, if it is indeed the case that a satisfactory solution to the experiential problem will require an agential conception of consciousness, the  QRF formalism as it is currently understood may not be providing us with the right kind of internal reference frame to address the experiential problem. Undoubtedly the formalism taking an important step in showing us how to extract a local description of physics out of the timeless picture - one way or another, we are clearly going to have to `zoom in' from a timeless, perspective-neutral picture to some internal perspective. But if the goal of zooming in is to understand temporal experience, we need to keep the self in the description when we adopt that internal perspective in order that we can sensibly model agency.  For as Ismael emphasizes, self-reference is an important aspect of our temporal experience, and yet neither a perspective-neutral global view nor the internal view from an observer's QRF permits any self-reference: a perspective-neutral picture does not admit any identification of a `self' at all, so it offers no way to identify what is actually experienced from within the large range of options represented by the wavefunction description, and meanwhile the internal view simply removes the self from the picture altogether. We need a different way of `zooming in' which does in some sense single out the observer as special, but which is not literally centered on the physical body of the relevant observer\footnote{Interestingly, there is an alternative approach to `quantum reference frames' within quantum information in which observers \emph{are} regarded as external to the relevant frames - the framework is used to study how observers can send reference frames to one another via exchange of physical information \citep{RevModPhys.79.555}. Perhaps this formalism might be relevant to the kind of reference frames described here. However, this formalism looks essentially like a version of the `perspective-neutral' view, since observers are present, but are described in a third-person, external way, so there does not appear to be any self-reference. In addition, it is not obvious how to relate an approach of this kind to the quantum-gravitational context, since the external description is not given from any perspective and yet still contains physical structures that according to the relational observables approach can only appear relative to a perspective, so possibly this kind of approach will only work in the context of non-relativistic quantum mechanics, or quantum mechanics with only weak gravity}.

  \section{Conclusion}

There appear to be  good reasons to think that in order to make  sense of temporal experience we must adopt an agential conception of consciousness, rather than the passive awareness approach common in philosphical work on this topic. And this suggests a potential shortcoming of the formalisms currently being used to study the emergence of time in the timeless context of quantum gravity.  In particular, the QRF formalism as it is currently used looks as if it is implicitly assuming a `passive awareness' model of consciousness: by switching into the reference frame associated with the physical body of an observer, we obtain what is referred to as 
a perspective, which  appears to be intended as a description of things that the agent might be aware of. But we cannot actually read awareness off an internal reference frame without first solving the experiential problem, and the passive awareness model does not look adequate to solve the experiential problem. We may need a more agential model of consciousness, and to do that we may have to employ some other kind of internal perspective; one which allows the possibility of  self-reference such that we can represent the consequences of possible future actions. 

Let me finish by noting that we may obtain some clues as to the kind of internal perspective needed by considering more carefully the nature of our own spatial and temporal experience.  The approach taken by the QRF formalism is known to psychologists of spatial representation as an `egocentric representation' \citep{https://doi.org/10.1002/brb3.1532}, in which objects are located by reference to one's own spatial location. But the way in which we  represent the world to ourselves in everyday life is not \emph{wholly} egocentric. Suppose I start  walking towards a table; if I were literally using myself as a reference frame, my representation of the world would represent that table as moving towards me, whereas in fact I understand this process as \emph{me} moving towards a table! Moreover, this feature is crucial to the practical utility of my representation. If I were to work in a reference frame centered on my body, then by definition I wouldn't be able  move about in that reference frame - if I wanted something that is sitting on the table, I would just have to wait and hope for the table to move towards me. So  in order that I can sensibly model the possibilities for my future action and use that modelling to decide what action I will take, I need to adopt a reference frame in which it is possible to describe myself as acting. As \cite{Rovellioriented} puts it, `\emph{our brain is essentially a machine that analyses the different possible futures that would follow if this or that course of action is taken. The main business our brain is involved in is not simply to predict the future given the past ... but to predict what would happen under different choices of behavior, namely to predict what would the effects of different interventions be.}' 

Yet at the same time, I am also not using  what the psychologists of spatial representation would call an `allocentric representation' \citep{https://doi.org/10.1002/brb3.1532} -  a third-person perspective-neutral model. For as Ismael's work highlights, in my modelling  my actions are afforded a special status, such that I cannot stabilize my knowledge of my own future actions. I need to be able to model my actions, but also still identify them as mine: in other words, I need self-reference. What I am doing is to some extent  what the psychologists of spatial representation call `spatial decentering' \citep{decentered} - the ability to adopt a spatial frame on an object from a location that isn't where the subject is located. But in this case there is no \emph{specific} location other than my own from which the reference frame is described, and the representation is not \emph{fully} decentered: I am still given a privileged place in my own representation, I am identified as the self, and my perceptions are all obtained at the physical location of my body. And yet I am still represented as moving around in a space, rather than having things move around relative to me.  So perhaps   the kind of reference frame we might need to properly model agency in a quantum context is something more like this - a reference frame which admits self-reference, and which is neither perspective-neutral nor perspectival in a crudely egocentric way.

\section{Acknowledgements}

Thanks to the organizers and participants of the `Schr\"{o}dinger's Cosmic Cat' workshop (Geneva, 2024) for very helpful discussions and comments.


\begin{thebibliography}{}

\bibitem[Adlam, 2022]{AdlamPW}
Adlam, E. (2022).
\newblock {Watching the Clocks: Interpreting the Page-Wootters Formalism and the Internal Quantum Reference Frame Programme}.
\newblock {\em Foundations of Physics}, 52(99).

\bibitem[Adlam, 2023]{adlam_2023}
Adlam, E. (2023).
\newblock The temporal asymmetry of influence is not statistical.
\newblock {\em Philosophy of Science}, page 1–18.

\bibitem[Albert, 2014]{Huckleberry}
Albert, D. (2014).
\newblock The sharpness of the distinction between the past and the future.
\newblock In Wilson, A., editor, {\em Chance and Temporal Asymmetry}. Oxford University Press.

\bibitem[Albert, 2000]{Albert2000-ALBTAC}
Albert, D.~Z. (2000).
\newblock {\em Time and Chance}.
\newblock Harvard University Press.

\bibitem[Andersen, 2014]{Andersen2014-ANDTDO-3}
Andersen, H. (2014).
\newblock {The Development of the Specious Present and James' Views on Temporal Experience}.
\newblock In Arstila, D. L.~V., editor, {\em Subjective Time: the philosophy, psychology, and neuroscience of temporality}, pages 25--42. MIT Press.

\bibitem[Arnold et~al., 2016]{decentered}
Arnold, G., Spence, C., and Auvray, M. (2016).
\newblock Taking someone else’s spatial perspective: Natural stance or effortful decentring?
\newblock {\em Cognition}, 148:27--33.

\bibitem[Bartlett et~al., 2007]{RevModPhys.79.555}
Bartlett, S.~D., Rudolph, T., and Spekkens, R.~W. (2007).
\newblock Reference frames, superselection rules, and quantum information.
\newblock {\em Rev. Mod. Phys.}, 79:555--609.

\bibitem[Block, 1995]{Block1995-BLOOAC}
Block, N. (1995).
\newblock On a confusion about a function of consciousness.
\newblock {\em Brain and Behavioral Sciences}, 18(2):227--247.

\bibitem[Boethius, 2016]{boethius2016consolation}
Boethius, A. (2016).
\newblock {\em The Consolation of Philosophy}.

\bibitem[Burge, 1979]{Burge1979-BURIAT-11}
Burge, T. (1979).
\newblock Individualism and the mental.
\newblock In Heil, J., editor, {\em Philosophy of Mind: A Guide and Anthology}. Oxford University Press.

\bibitem[Callender, 2017]{Callender2017-CALWMT}
Callender, C. (2017).
\newblock {\em What Makes Time Special?}
\newblock Oxford University Press, Oxford.

\bibitem[Carette et~al., 2023]{carette2023operational}
Carette, T., Głowacki, J., and Loveridge, L. (2023).
\newblock Operational quantum reference frame transformations.

\bibitem[Castro-Ruiz et~al., 2020]{2020qctloe}
Castro-Ruiz, E., Giacomini, F., Belenchia, A., and Brukner,  . (2020).
\newblock Quantum clocks and the temporal localisability of events in the presence of gravitating quantum systems.
\newblock {\em Nature Communications}, 11(1).

\bibitem[Chiang, 2010]{chiang2010stories}
Chiang, T. (2010).
\newblock {\em Stories of Your Life and Others}.
\newblock Knopf Doubleday Publishing Group.

\bibitem[Curiel, 2020]{curiel2020schematizing}
Curiel, E. (2020).
\newblock Schematizing the observer and the epistemic content of theories.

\bibitem[Davidson, 1963]{60a9dd9a-e48a-3fcf-87dd-d5c158406fc6}
Davidson, D. (1963).
\newblock Actions, reasons, and causes.
\newblock {\em The Journal of Philosophy}, 60(23):685--700.

\bibitem[de~la Hamette et~al., 2021]{delahamette2021perspectiveneutral}
de~la Hamette, A.-C., Galley, T.~D., Hoehn, P.~A., Loveridge, L., and Mueller, M.~P. (2021).
\newblock Perspective-neutral approach to quantum frame covariance for general symmetry groups.

\bibitem[Earman, 2002]{Earman2002-EARTMM}
Earman, J. (2002).
\newblock {Thoroughly Modern McTaggart: Or, What McTaggart Would Have Said If He Had Read the General Theory of Relativity}.
\newblock {\em Philosophers' Imprint}, 2:1--28.

\bibitem[Einstein, 1918]{Einsteindialog}
Einstein, A. (1918).
\newblock Dialog über einwände gegen die relativitätstheorie.
\newblock page 697–702.

\bibitem[Fernandez-Baizan et~al., 2020]{https://doi.org/10.1002/brb3.1532}
Fernandez-Baizan, C., Nuñez, P., Arias, J.~L., and Mendez, M. (2020).
\newblock Egocentric and allocentric spatial memory in typically developed children: Is spatial memory associated with visuospatial skills, behavior, and cortisol?
\newblock {\em Brain and Behavior}, 10(5):e01532.

\bibitem[Fields et~al., 2022]{Fields_2022}
Fields, C., Glazebrook, J.~F., and Levin, M. (2022).
\newblock Neurons as hierarchies of quantum reference frames.
\newblock {\em Biosystems}, 219:104714.

\bibitem[Gary and Giddings, 2007]{Gary_2007}
Gary, M. and Giddings, S.~B. (2007).
\newblock Relational observables in 2d quantum gravity.
\newblock {\em Physical Review D}, 75(10).

\bibitem[Giacomini and Brukner, 2022]{giacomini2021quantum}
Giacomini, F. and Brukner, {\v{C} }. (2022).
\newblock Quantum superposition of spacetimes obeys einstein{\textquotesingle}s equivalence principle.
\newblock {\em {AVS} Quantum Science}, 4(1):015601.

\bibitem[Giacomini et~al., 2019]{article3}
Giacomini, F., Castro-Ruiz, E., and Brukner, C. (2019).
\newblock Quantum mechanics and the covariance of physical laws in quantum reference frames.
\newblock {\em Nature Communications}, 10.

\bibitem[Healey, 2002]{Healey2002-HEACPC}
Healey, R. (2002).
\newblock Can physics coherently deny the reality of time?
\newblock {\em Royal Institute of Philosophy Supplement}, 50:293--.

\bibitem[Höhn and Vanrietvelde, 2020]{2020htsbrqc}
Höhn, P.~A. and Vanrietvelde, A. (2020).
\newblock How to switch between relational quantum clocks.
\newblock {\em New Journal of Physics}, 22(12):123048.

\bibitem[Ismael, 2023]{doi:10.1080/24740500.2022.2155200}
Ismael, J. (2023).
\newblock The open universe: Totality, self-reference and time.
\newblock {\em Australasian Philosophical Review}, 0(0):1--16.

\bibitem[Kirk, 2023]{sep-zombies}
Kirk, R. (2023).
\newblock {Zombies}.
\newblock In Zalta, E.~N. and Nodelman, U., editors, {\em The {Stanford} Encyclopedia of Philosophy}. Metaphysics Research Lab, Stanford University, {F}all 2023 edition.

\bibitem[Kutach, 2013]{kutach2013causation}
Kutach, D. (2013).
\newblock {\em Causation and Its Basis in Fundamental Physics}.
\newblock Oxford Studies in Philosophy of Science. OUP USA.

\bibitem[Lam and Wüthrich, 2023]{Lam_2023}
Lam, V. and Wüthrich, C. (2023).
\newblock Laws beyond spacetime.
\newblock {\em Synthese}, 202(3).

\bibitem[Maudlin, 2002]{Maudlin2002-MAUTMM}
Maudlin, T. (2002).
\newblock Thoroughly muddled mctaggart: Or, how to abuse gauge freedom to create metaphysical monostrosities.
\newblock {\em Philosophers' Imprint}, 2:1--23.

\bibitem[McKenna, 2023]{McKenna2023-MCKAAT-10}
McKenna, C.~A. (2023).
\newblock Agency and the successive structure of time-consciousness.
\newblock {\em Erkenntnis}, 88(5):2013--2034.

\bibitem[Pautz, 2019]{Pautz2019-PAUWIT-2}
Pautz, A. (2019).
\newblock What is the integrated information theory of consciousness?
\newblock {\em Journal of Consciousness Studies}, 26(1-2):1--2.

\bibitem[Putnam, 1967]{Putnam}
Putnam, H. (1967).
\newblock Time and physical geometry.
\newblock {\em Journal of Philosophy}, 64(8):240--247.

\bibitem[Putnam, 1975]{Putnam1975-PUTTMO}
Putnam, H. (1975).
\newblock {The Meaning of `Meaning'}.
\newblock {\em Minnesota Studies in the Philosophy of Science}, 7:131--193.

\bibitem[Rickles, 2008]{pittphilsci4223}
Rickles, D. (2008).
\newblock Who's afraid of background independence?
\newblock Later published version in D. Dieks (ed.), The Ontology of Spacetime II (pp. 133-152). Elsevier, 2008.

\bibitem[Rovelli, 2002]{Rovelli_2002}
Rovelli, C. (2002).
\newblock Partial observables.
\newblock {\em Physical Review D}, 65(12).

\bibitem[Rovelli, 2014]{Rovelli_2014}
Rovelli, C. (2014).
\newblock Why gauge?
\newblock {\em Foundations of Physics}, 44(1):91–104.

\bibitem[Rovelli, 2021]{rovelli2021layers}
Rovelli, C. (2021).
\newblock The layers that build up the notion of time.

\bibitem[{Rovelli}, 2023]{Rovellioriented}
{Rovelli}, C. (2023).
\newblock {How Oriented Causation Is Rooted into Thermodynamics}.
\newblock 1(1):11.

\bibitem[Rovelli and Vidotto, 2022]{rovelli2022philosophical}
Rovelli, C. and Vidotto, F. (2022).
\newblock Philosophical foundations of loop quantum gravity.

\bibitem[Shapiro and Spaulding, 2024]{sep-embodied-cognition}
Shapiro, L. and Spaulding, S. (2024).
\newblock {Embodied Cognition}.
\newblock In Zalta, E.~N. and Nodelman, U., editors, {\em The {Stanford} Encyclopedia of Philosophy}. Metaphysics Research Lab, Stanford University, {F}all 2024 edition.

\bibitem[Silberstein et~al., 2018]{10.1093/oso/9780198807087.003.0003}
Silberstein, M., Stuckey, W., and McDevitt, T. (2018).
\newblock {53The Block Universe from Special Relativity}.
\newblock In {\em {Beyond the Dynamical Universe: Unifying Block Universe Physics and Time as Experienced}}. Oxford University Press.

\bibitem[Smith and Ahmadi, 2020]{2020cqtd}
Smith, A. R.~H. and Ahmadi, M. (2020).
\newblock Quantum clocks observe classical and quantum time dilation.
\newblock {\em Nature Communications}, 11(1).

\bibitem[Stein, 1994]{Stein}
Stein, H. (1994).
\newblock {Some Reflections on the Structure of our Knowledge in Physics}.
\newblock In D.~Prawitz, B.~S. and Westerst\o{a}hl, D., editors, {\em {Logic, Methodology and Philosophy of Science}}, pages 633--655. Elsevier Science B.V.

\bibitem[Th\'{e}bault, 2019]{pittphilsci15795}
Th\'{e}bault, K. P.~Y. (2019).
\newblock The problem of time.
\newblock Forthcoming in the `Routledge Companion to the Philosophy of Physics' edited by Eleanor Knox and Alastair Wilson.

\bibitem[Tononi et~al., 2016]{tononi}
Tononi, G., Boly, M., Massimini, M., and Koch, C. (2016).
\newblock Integrated information theory: From consciousness to its physical substrate.
\newblock {\em Nature Reviews Neuroscience}, 17.

\bibitem[Vanrietvelde et~al., 2020]{2020acop}
Vanrietvelde, A., Hoehn, P.~A., Giacomini, F., and Castro-Ruiz, E. (2020).
\newblock A change of perspective: switching quantum reference frames via a perspective-neutral framework.
\newblock {\em Quantum}, 4:225.

\bibitem[Vonnegut, 1998]{vonnegut1998sirens}
Vonnegut, K. (1998).
\newblock {\em The Sirens of Titan: A Novel}.
\newblock A Delta Book. Random House Publishing Group.

\bibitem[Wallace, 2002]{Wallacenew}
Wallace, D. (2002).
\newblock In Brading, K. and Castellani, E., editors, {\em Symmetries in Physics: Philosophical Reflections}, pages 163--173. Cambridge University Press.

\bibitem[Young, 2022]{Young2022AgentsOC}
Young, N. (2022).
\newblock Agents of change: temporal flow and feeling oneself act.
\newblock {\em Philosophical Studies}, 179:2619 -- 2637.

\end{thebibliography}
\end{document}